\documentstyle[prl,aps,multicol,epsf]{revtex}
\begin{document}
\draft
\title{Theory of magnetic excitations in the Kondo lattice}
\author{R. Eder}
\address{Institut f\"ur Theoretische Physik, Universit\"at W\"urzburg,
Am Hubland,  97074 W\"urzburg, Germany
}
\date{\today}
\maketitle
\begin{abstract}
We present a theory for the spin excitations of the Kondo lattice. We derive
an effective Hamiltonian, which describes Fermionic spin 1/2 charge 
fluctuations interacting with Bosonic triplet spin fluctuations. Evaluating 
the polarization bubble in the magnon Green's function with the free charge 
excitation's Green's function gives a quantitative description of the `spin 
gap' and `charge gap' as obtained from numerical
calculations for the Kondo insulator.
\end{abstract}
\begin{figure}
\epsfxsize=6.1cm
\vspace{0.5cm}
\hspace{1.0cm}\epsffile{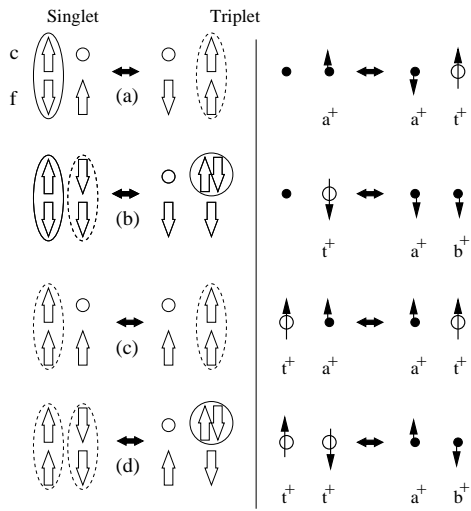}
\vspace{0.5cm}
\narrowtext
\caption[]{Different hopping processes which involve triplet cells
(left panel) and their representation in terms of  book-keeping Fermions
and Bosons (right panel). The transition is always
accomplished by the hopping of a physical $\uparrow$-electron
from left to right. Spin and site labels on the
book-keeping particles in the left panel are suppressed for simplicity.}
\label{fig1} 
\end{figure}
\begin{figure}
\epsfxsize=6.1cm
\vspace{0.5cm}
\hspace{0.5cm}\epsffile{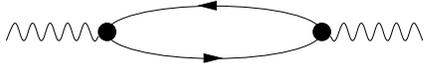}
\vspace{0.5cm}
\narrowtext
\caption[]{The self-energy $\Sigma(\bbox{q},\omega)$
used for the calculation of the
triplet dispersion (the wiggly triplet legs to the right and
left must be removed).}
\label{fig2} 
\end{figure}
\begin{figure}
\epsfxsize=6.1cm
\vspace{0.5cm}
\hspace{0.5cm}\epsffile{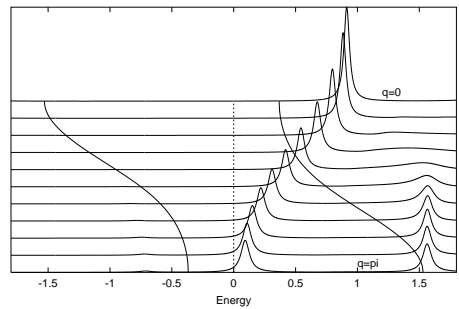}
\vspace{0.5cm}
\narrowtext
\caption[]{Triplet spectral density $A(\bbox{q},\omega)$
for the 1D Kondo lattice
with $J/t$$=$$1.0$, as a function of frequency and momentum. The
Lorentzian broadening $\epsilon$$=$$0.05$.}
\label{fig3} 
\end{figure}
\begin{figure}
\epsfxsize=6.1cm
\vspace{0.5cm}
\hspace{0.5cm}\epsffile{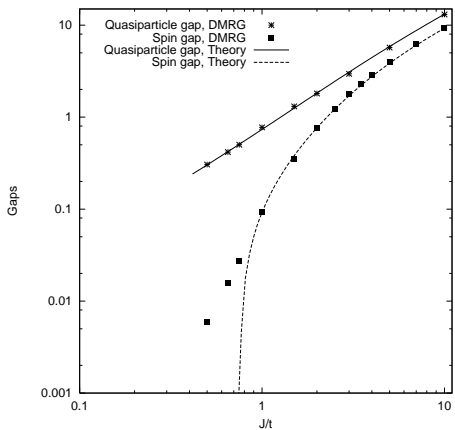}
\vspace{0.5cm}
\narrowtext
\caption[]{Comparison of spin gap and quasiparticle gap
in the 1D Kondo lattice 
as obtained by Yu and White\cite{YuWhite} from DMRG calculations
on a chain with $24$ sites to the results of the present theory.}
\label{fig4} 
\end{figure}

\begin{multicols}{2}
The Kondo lattice model is a generic model of strongly correlated electrons.
It may be applicable, with some variations, to a wide variety of materials, 
most notably the
Heavy Fermion metals\cite{Bickers,Fulde} and Kondo insulators\cite{Fisk}.
It is the purpose of the present manuscript
to present a conceptually very simple theory for the
spin and charge excitations of this model, and their
mutual interplay. The basic idea is, in the spirit of the cell-perturbation
theory developed by Jefferson and co-workers\cite{Jefferson},
to split the systems up into sub-units which can be
solved exactly. Coupling the sub-units together the leads
to a simple effective Hamiltonian. As has been shown
in Refs.\cite{eos} for the one particle-spectra and as will be shown
in the present work for the spin excitations, solving this Hamiltonian
with the simplest approximations possible produces already
quite satisfactory results. An approach which is
similar in spirit has been proposed for the Kondo lattice by by 
P\'erez-Conde {\em et al.}\cite{PerezConde}
and for the description of
$\pi$-electrons in conjugated polymers by Pleutin\cite{Pleutin}.\\
The Hamiltonian considered in this work reads
\begin{equation}
H_{sc} =
\sum_{\bbox{k},\sigma} \epsilon_{\bbox{k}}\;
c_{\bbox{k},\sigma}^\dagger c_{\bbox{k},\sigma}^{} 
+ J \sum_n \bbox{S}_{n,c} \cdot \bbox{S}_{n,f}.
\label{kondo2}
\end{equation}
Here $\epsilon_{\bbox{k}}$$=$$\sum_{n}
exp(i\bbox{k} \cdot(\bbox{R}_n - \bbox{R}_m))\; t_{mn}$
is the Fourier transform of the
hopping integral $t_{nm}$ for the $c$ electrons,
and $\bbox{S}_{n,c}$ $(\bbox{S}_{n,f})$ denotes the
spin operator for conduction electrons ($f$-electrons)
in cell $n$. A unit cell contains one $c$-orbital and one
localized spin.\\
The physical parameter regime is
$J$$<$$max|t_{nm}|$. It was shown previously\cite{eos}, however,
that an `expansion' around the limiting case $t_{mn}$$=$$0$,
where the ground state for one conduction electron/unit cell
is simply a product of single-cell singlets, gives in fact a surprisingly
good description even in the physical parameter regime.
The basic idea is a mapping between a suitably restricted Hilbert space
and states of effective Fermions which basically stand for
charge fluctuations in the singlet background. More precisely,
when a single cell is in
the two-electron singlet state, we consider this cell
to be empty. If the cell $n$ is occupied only by the
$f$-spin we consider it occupied by a hole-like Fermion,
created by $a_{n,\sigma}^\dagger$, if it is
occupied by three electrons we model this
by the presence of an electron-like Fermion, created by
 $b_{n,\sigma}^\dagger$. So far our `translation table' thus reads:
\begin{eqnarray}
|vac\rangle &\rightarrow& \frac{1}{\sqrt{2}} 
(c_{n,\uparrow}^\dagger f_{n,\downarrow}^\dagger
- c_{n,\downarrow}^\dagger f_{n,\uparrow}^\dagger)|0\rangle,
\nonumber \\
a_{n,\sigma}^\dagger|vac\rangle
&\rightarrow& f_{n,\sigma}^\dagger |0\rangle,
\nonumber \\
b_{n,\sigma}^\dagger|vac\rangle
&\rightarrow& c_{n,\uparrow}^\dagger c_{n,\downarrow}^\dagger 
f_{n,\sigma}^\dagger |0\rangle.
\label{trans1}
\end{eqnarray}
In the reduced Hilbert space which is built up from
product states of single cell states of the type (\ref{trans1}),
the Hamiltonian (\ref{kondo2}) takes the form\cite{eos}
\begin{eqnarray}
H_{eff} &=& \frac{1}{2}\sum_{\bbox{k},\sigma}
[\;(-\epsilon_{\bbox{k}}+ \frac{3J}{2}) a_{\bbox{k},\sigma}^\dagger
a_{\bbox{k},\sigma}^{} +
(\epsilon_{\bbox{k}}+\frac{3J}{2} )
b_{\bbox{k},\sigma}^\dagger b_{\bbox{k},\sigma}^{} \;]
\nonumber \\
&-& \frac{1}{2}\sum_{\bbox{k},\sigma}
sign(\sigma)\; \epsilon_{\bbox{k}} \;
(b_{\bbox{k},\sigma}^\dagger a_{-\bbox{k},\bar{\sigma}}^\dagger
 + H.c.).
\label{stcham}
\end{eqnarray}
The Fermions have to obey a hard-core constraint,
i.e. no cell may be occupied by more than one Fermion. This constraint
is necessary to make sure that the state of any given cell be unique.
If to simplest approximation we relax this
constraint, the Hamiltonian (\ref{stcham}) can be solved
by Bogoliubov transformation. Introducing $\Delta$$=$$3J/2$, the
quasiparticle dispersion\cite{eos} is
\begin{equation}
E_\pm(\bbox{k}) = (1/2)\;[\; \epsilon_{\bbox{k}} \pm
\sqrt{ \epsilon_{\bbox{k}}^2+ \Delta^2  }\;],
\label{scdisp}
\end{equation}
with the resulting quasiparticles
\begin{eqnarray}
\gamma_{\bbox{k},-,\sigma} &=& \;\;u_{\bbox{k}} b_{\bbox{k},\sigma}^{}
+ sign(\sigma)v_{\bbox{k}} a_{-\bbox{k},\bar{\sigma}}^\dagger,
\nonumber \\
\gamma_{\bbox{k},+,\sigma} &=& -v_{\bbox{k}} b_{\bbox{k},\sigma}^{}
+  sign(\sigma) u_{\bbox{k}} a_{-\bbox{k},\bar{\sigma}}^\dagger,
\label{ansatz}
\end{eqnarray}
The choice of signs ensures that e.g. $[S^+,\gamma_{\uparrow}]
$$=-$$\gamma_{\downarrow}$.\\
The Hamiltonian (\ref{stcham}) operates only within a reduced
Hilbert space. Any states where a unit cell
with two electrons (one $f$ and one $c$-electron)
is in the {\em triplet} state are projected
out. This restriction makes the propagation
of the charge fluctuations $\gamma^\dagger$ completely coherent,
and is justified to some degree by the quite good
agreement of the resulting
single-particle spectra with Lanczos diagonalization\cite{eos} - 
but ultimately it is of course an uncontrolled approximation.
Therefore we now want to enlarge the Hilbert space 
to its full size by introducing an additional effective particle
$\bbox{t}_i$, whose presence in a given cell $i$ implies that this cell
is occupied by two electrons in the triplet state.
Since the three components of
a triplet are isomorphic to a 3-vector, $\bbox{t}_i$
must be a vector particle. 
We thus enlarge the `translation table' (\ref{trans1}) by
\begin{eqnarray}
\bbox{t}_{n}^\dagger |vac \rangle
\rightarrow 2i\sqrt{2}\;
c_{n,\tau}^\dagger (\bbox{\sigma}\sigma^y)_{\tau,\tau'} 
 f_{n,\tau'}^\dagger |0\rangle,
\end{eqnarray}
where $\bbox{\sigma}$ is the vector of Pauli matrices
(with $\sigma^2$$=$$3/4$)
and repeated spin indices $\tau,\tau'$ are summed over.\\
The quantum numbers of the $\bbox{t}^\dagger$ particle
are spin $1$ and charge $0$, whence it must be a Boson.
As was the case for the Fermions, the Bosons have to obey a
mutual exclusion principle, i.e.
no two particles, of either Fermionic or Bosonic nature,
must occupy the same site.\\
Having introduced the $\bbox{t}^\dagger$ Bosons to mark
spin excited cells, we proceed to derive the
Hamiltonian which describes their interaction with the
charged excitations $a^\dagger$ and $b^\dagger$.
Figure \ref{fig1} gives a survey of the different interaction 
\noindent 
processes that are possible. The form of the respective terms in the 
Hamiltonian can be inferred from the
requirement of Hermiticity, spin rotation and time reversal invariance.
The first type of interaction process we consider involves
two Fermion operators and one Boson operator.
They correspond first to the emission (or absorption) of a 
triplet as a charge fluctuation hops from one cell
into the other (see Figure \ref{fig1}a). Another process of this type
is the pair creation of one electron-like and
one hole-like charge fluctuation in which a triplet is
annihilated (see Figure \ref{fig1}b).
The only way to construct a spin scalar from two spinors
(say: $a^\dagger$ and $a$)
and one vector ($\bbox{t}^\dagger$) is to
first couple the spinors into a vector:
\begin{eqnarray}
\bbox{S}_{mn}^a &=& 
\;\;\;a_{m,\tau}^\dagger
\bbox{\sigma}_{\tau,\tau'} a_{n,\tau'}^{},
\nonumber \\
\bbox{S}_{mn}^b &=&\; \;\;b_{m,\tau}^\dagger
\bbox{\sigma}_{\tau,\tau'} b_{n,\tau'}^{},
\nonumber \\
\bbox{S}_{mn}^+ &=& 2i b_{m,\tau}^\dagger
(\bbox{\sigma} \sigma^y)_{\tau,\tau'} a_{n,\tau'}^{\dagger},
\label{vector}
\end{eqnarray}
and then form its scalar product with the vector $\bbox{t}^\dagger$.
Due to the product nature of the basis states,
the matrix elements can be evaluated by considering
just two cells $m$ and $n$ connected by the
hopping integral $t_{mn}$\cite{eos}. The result is:
\begin{eqnarray}
H_1 &=& \sum_{m,n}\;t_{mn}\; [ \;
\bbox{t}_m^{} \cdot \; (\;\bbox{S}_{mn}^a 
-  \bbox{S}_{mn}^b\;)
\nonumber \\
&+& (\bbox{t}_m^{} - \bbox{t}_n^{})\cdot \; \bbox{S}_{mn}^+   
\;)\;]\; + H.c.
\end{eqnarray}
Next, we consider processes involving two Fermions
and two Bosons.
Examples are shown in Figure \ref{fig1}c, where a triplet
and a charge fluctuation exchange their position,
and in Figure \ref{fig1}d, where two triplets
are annihilated in the pair creation of one electron-like
and one hole-like charge fluctuation.
There are two ways to from a spin scalar from two
spinors and two vectors: first,
one can contract both the spinors and the vectors separately
into scalars and multiply these. This term describes
processes where the Fermion and the Boson exchange their position
without exchanging spin. Alternatively, one can
combine the two spinors into a vector following (\ref{vector}),
and form the wedge product of the three resulting vectors.
We thus find the additional terms
\begin{eqnarray}
H_2 &=&  \sum_{m,n} \;\frac{t_{mn}}{2} \;
\bbox{t}_m^\dagger \cdot \bbox{t}_n^{}
\sum_\sigma (\;b_{n,\sigma}^\dagger b_{m,\sigma}^{}
- a_{m,\sigma}^\dagger a_{n,\sigma}^{}\;)\;+
\nonumber \\
\sum_{m,n}&[&\frac{t_{mn}}{2}
\bbox{t}_m^\dagger \cdot \bbox{t}_n^\dagger
\sum_\sigma sign(\sigma)\;b_{n,\bar{\sigma}}^{} a_{m,\sigma}^{}\;
 + H.c.\;]
\nonumber \\
H_3 &=& i \sum_{m,n} \;t_{mn}\;
(\;\bbox{S}_{mn}^a -\bbox{S}_{mn}^b\;) \cdot 
(\; \bbox{t}_n^\dagger \times \bbox{t}_m^{} \;)
\nonumber \\
&+&i \sum_{m,n} \;[\;t_{mn}\;
\bbox{S}_{mn}^+\cdot 
(\; \bbox{t}_n^\dagger \times \bbox{t}_m^{\dagger} \;)\;
+ H.c.\;]
\end{eqnarray}
The terms $H_1$, $H_2$ and $H_3$ combined describe all 
possible interactions of the vector Bosons with the
Fermionic charge fluctuations.
The only term still missing to complete the Hamiltonian is
the `energy of formation' of the triplets:
$H_4$$=$$ J \sum_n \bbox{t}_n^\dagger \cdot \bbox{t}_n^{}$.\\
Having written down the Hamiltonian we proceed to
investigate the physics resulting from it. For simplicity
we restrict ourselves in this paper to the case of half-filling,
i.e. the `Kondo insulator'.
As our overall guideline we try to keep the
approximations involved as simple as possible and consequently
neglect the $4$-particle vertices $H_2$ and $H_3$.
This leaves us with $H_1$, which describes the propagation
of a triplet by an RKKY-like process:
the triplet may be converted into an electron-hole pair, which
may later recombine at another site.
Alternatively, it may be absorbed by an electron-like or hole-like
charge fluctuation, and be re-emitted at a different site.
After Fourier and Bogoliubov transformation of $H_1$:
\begin{eqnarray*}
H_1 =\frac{1}{\sqrt{N}}&&
\sum_{\bbox{q},\bbox{k}} \sum_{\mu,\nu=\pm}
M^{\mu,\nu}_{\bbox{k},\bbox{q}}\; t_{\bbox{q}}^\dagger \cdot
\gamma_{\bbox{k},\mu,\tau}^\dagger \bbox{\sigma}_{\tau,\tau'}
\gamma_{\bbox{k}+\bbox{q},\nu,\tau'}^{},
\nonumber \\
M^{-,-}_{\bbox{k},\bbox{q}} &=&
\epsilon_{\bbox{k}+\bbox{q}} v_{\bbox{k}} (v_{\bbox{k}+\bbox{q}}
-u_{\bbox{k}+\bbox{q}}) - \epsilon_{\bbox{k}}
u_{\bbox{k}+\bbox{q}} (u_{\bbox{k}} -v_{\bbox{k}}),
\nonumber \\
M^{-,+}_{\bbox{k},\bbox{q}} &=&
\epsilon_{\bbox{k}+\bbox{q}} v_{\bbox{k}} (v_{\bbox{k}+\bbox{q}}
+u_{\bbox{k}+\bbox{q}}) + \epsilon_{\bbox{k}}
v_{\bbox{k}+\bbox{q}} (u_{\bbox{k}} -v_{\bbox{k}}),
\nonumber \\
M^{+,-}_{\bbox{k},\bbox{q}} &=&
\epsilon_{\bbox{k}+\bbox{q}} u_{\bbox{k}} (v_{\bbox{k}+\bbox{q}}
-u_{\bbox{k}+\bbox{q}}) + \epsilon_{\bbox{k}}
u_{\bbox{k}+\bbox{q}} (u_{\bbox{k}} +v_{\bbox{k}}),
\nonumber \\
M^{+,+}_{\bbox{k},\bbox{q}} &=&
\epsilon_{\bbox{k}+\bbox{q}} u_{\bbox{k}} (v_{\bbox{k}+\bbox{q}}
+u_{\bbox{k}+\bbox{q}}) - \epsilon_{\bbox{k}}
v_{\bbox{k}+\bbox{q}} (u_{\bbox{k}} +v_{\bbox{k}}),
\end{eqnarray*}
we introduce the triplet Green's function
(where $\alpha$ may be $x,y$ or $z$):
\begin{eqnarray}
D(\bbox{q},\omega) &=&
-i \langle T t_{\bbox{q},\alpha}^\dagger t_{\bbox{q},\alpha}^{} \rangle
\nonumber \\
&=& \lim_{\epsilon \rightarrow 0^+}
\frac{1}{\omega - J - \Sigma(\bbox{q},\omega) + i\epsilon}.
\end{eqnarray}
For the self-energy $\Sigma(\bbox{q},\omega)$ we choose the simplest diagram 
possible, i.e. the polarization bubble in Figure \ref{fig2}.
\noindent 
The result is:
\begin{eqnarray}
\Sigma(\bbox{q},\omega)&=&
\frac{-i}{2N} 
\sum_{\bbox{k}}\sum_{\mu,\nu=\pm} |M^{\mu,\nu}_{\bbox{k},\bbox{q}}|^2
\nonumber \\
&\;&\;\;\;\;\;\;\;
\int_{-\infty}^{\infty} 
\frac{d\omega'}{2\pi} G_\nu(\bbox{k}+\bbox{q},\omega'+\omega)
G_\mu(\bbox{k},\omega'),
\nonumber \\
G_\mu(\bbox{k},\omega')&=& 
\frac{\delta_{\mu,-}}{\omega - E_-(\bbox{k}) -i\epsilon}
+ \frac{\delta_{\mu,+}}{\omega - E_+(\bbox{k}) +i\epsilon}
\end{eqnarray}
(we are using the free Green's function $G_\mu(\bbox{k},\omega')$ for the
charge fluctuations $\gamma^\dagger$).\\
To describe the low energy scales of the Kondo lattice at least
approximately, we still have to deal with the constraints.
We thereby follow Gopalan {\em et al.}\cite{Gopalan} 
and use a simple mean-field like
renormalization of the hopping integral. We consider
the vacuum state, where each site is occupied by a singlet,
as a condensate of an additional Bosonic
particle, $s^\dagger_n$, which stands for
to the two-electron singlet state. Then, in the Fermionic Hamiltonian
(\ref{stcham}) and in th eterm $H_1$ we have to replace
\begin{eqnarray*}
b_{m,\sigma}^\dagger a_{n,\sigma}^\dagger
&\rightarrow& (b_{m,\sigma}^\dagger s_m)\;(a_{n,\sigma}^\dagger s_n),
\nonumber \\
a_{m,\sigma}^\dagger a_{n,\sigma}
&\rightarrow& (a_{m,\sigma}^\dagger s_m)\;(s_n^\dagger a_{n,\sigma}),
\nonumber \\
\bbox{t}_m^{} \cdot \; \bbox{S}_{mn}^a 
&\rightarrow&
s_n^\dagger \bbox{t}_m^{} \cdot \; \bbox{S}_{mn}^a
\end{eqnarray*}
(and analogously for the other terms).
The constraint takes the form
\[
s_n^\dagger s_n^{}
+ \sum_\sigma (a_{n,\sigma}^\dagger a_{n,\sigma}^{} +
b_{n,\sigma}^\dagger b_{n,\sigma}^{})
+ \bbox{t}_n^\dagger \cdot \bbox{t}_n^{} = 1.
\]
We now assume that the $s^\dagger$ singlet is condensed\cite{Gopalan},
so that the respective operators can be replaced by the real number $s$
which is determined from
\begin{equation}
s^2 = 1 -\sum_\sigma (a_{n,\sigma}^\dagger a_{n,\sigma}^{} +
b_{n,\sigma}^\dagger b_{n,\sigma}^{})
\label{mf}
\end{equation}
(this means that we neglect the density of triplets, 
$\sum_n \bbox{t}_n^\dagger \cdot \bbox{t}_n^{}$, which is a good approximation
for large and moderate values of $J$).
All in all, in the Hamiltonian (\ref{stcham})
we then have to replace throughout $\epsilon_{\bbox{k}}
\rightarrow s^2\epsilon_{\bbox{k}}$, whereas $H_1$ changes according to
$H_1 \rightarrow s H_1$. \\
Figure \ref{fig3} then shows the calculated triplet spectral density,
$A(\bbox{q},\omega)$$=$$\Im (1/\pi) D(\bbox{q},\omega)$.
This shows one dominant magnetic excitation,
wich has quite a substantial dispersion of order
$J$. Such a sharp low-energy mode which appears only in the
spin correlation function has also been observed in
Quantum Monte-Carlo simulations at low temperatures\cite{groebereder}.
Near $q$$=$$\pi$ the dispersion takes its minimum and thus
defines the `spin gap', $\Delta_s$, i.e. the lowest
energy required to create a magnetic excitation
out of the ground state.
Also shown in Figure \ref{fig3} is the dispersion of the two
single particle bands, $E_\pm(k)$. Were 
\noindent 
it not for the low-energy triplet mode,
the lowest possible excitation would
correspond to exciting one Fermion
from the occupied band into the unoccupied one. 
The energy required for this
is the so-called quasiparticle gap
$\Delta_{QP}$$=$$E_-(k$$=$$\pi)$$-$$E_+(k$$=$$0)$. 
If the interaction between the particle-hole pair created in this
way is small , this energy should also correspond to the
so-called charge gap $\Delta_c$, the energy of the lowest excitation
observable in the density correlation function.
In fact, density matrix
renormalization group (DMRG) calculations by Yu and White\cite{YuWhite}
have shown that this neglect of interaction is quite a good aproximation.
To check our theory, we can thus compare the
quasiparticle gap $\Delta_{QP}\approx\Delta_{c}$ and the
spin gap, $\Delta_s$, to the
results of Yu and White. This is done in Figure \ref{fig4}
for different values of the Kondo exchange $J$. 
The gaps $\Delta_{QP}$ and $\Delta_s$
differ strongly - as is to be expected because they correspond
to completely different types of excitations, i.e. the transfer
of a charge fluctuation from the lower to the
upper band on one hand, and the propagating triplet on the other. For
$J/t\ge1$ the agreement between theory and numerics is quantitative -
this demonstrates that even our very simple and 
\noindent 
non-selfconsistent calculation of the Green's function apparently captures 
the essential physics. For small values of $J$ our present approximation
is not to be trusted any longer. The main reason is
that for small $J/t$
the probability for charge fluctuations becomes very large
($J$ sets the energy scale for the pair-breaking energy) whence
our mean-field approximation (\ref{mf})
predicts a quite dramatic renormalization of
$\epsilon_{\bbox{k}}$ - such a renormalization, however,
is not observed in Lanczos calculations\cite{eos}.
Surprisingly enough, Figure \ref{fig4} shows that
despite this problem the estimate for the quasiparticle gap
remains accurate even for small $J/t$ - this agreement, however, 
may as well be fortuitious.\\
In summary, we have derived a theory for the dynamics of
spin and charge excitations in the Kondo lattice and their interaction.
The `purely Fermionic' version, i.e. excluding the spin excitations,
has recently been shown to give quite good results when compared
to Lanczos diagonalization, as a matter of fact for a wide variety of different
versions of the Kondo lattice\cite{eos}. Augmenting the formalism by
Bosonic spin excitations and using the simplest approximations possible,
i.e.  a mean-field approximation to deal with the constraint
and a lowest order evaluation of the Fermionic polarization bubble,
we could obtain also a quite satisfactory description of the 
spin dynamics. In particular, at least for moderate values of
the Kondo exchange $J$, a quantitative description of the spin and charge gap
as inferred from DMRG calculations was possible.
This gives reason for some optimism to create, mainly by
using more sophisticated methods to deal with the 
constraint\cite{Kotov} in the case of small $J/t$, a theory which is
capable of describing the Kondo lattice from the largest energy scales
(i.e. those of the Hubbard bands) down to the smallest ones (i.e. the
spin gap). It should also be noted that the formalism 
is independent of dimensionality or lattice geometry.
It involves comparatively little
numerical effort and is therefore not restricted either to the somewhat
`toy model type' systems considered so far. In fact, both the
quasiparticle dispersion and the self-energy for the spin excitations
involve only the unhybridized conduction band dispersion
$\epsilon_{\bbox{k}}$. Using here the results from a `frozen core'
LDA band structure calculation may allow to apply the present formalism
rather directly to quite realistic systems.\\
I would like to thank Dr. C. C. Yu and Dr. S. R. White for providing
me with the data from Ref. \cite{YuWhite}.

 
\end{multicols}
\end{document}